# Toxicity Detection in Drug Candidates using Simplified Molecular-Input Line-Entry System

Nath Mriganka
B. Tech, ECE
National Institute of Technology, Silchar
India

Goswami Subhasish
B. Tech, CSE
Tezpur University
India

## ABSTRACT
The need for analysis of toxicity in new drug candidates and the requirement of doing it fast have asked the consideration of scientists towards the use of artificial intelligence tools to examine toxicity levels and to develop models to a degree where they can be used commercially to measure toxicity levels efficiently in upcoming drugs. Artificial Intelligence based models can be used to predict the toxic nature of a chemical using Quantitative Structure–Activity Relationship techniques. Convolutional Neural Network models have demonstrated great outcomes in predicting the qualitative analysis of chemicals in order to determine the toxicity. This paper goes for the study of Simplified Molecular-Input Line-Entry System (SMILES) as a parameter to develop Long-short term memory (LSTM) based models in order to examine the toxicity of a molecule and the degree to which the need can be fulfilled for practical use alongside its future outlooks for the purpose of real world applications.

## Keywords
Artificial Intelligence, Convolutional Neural Network, Simplified Molecular-Input Line-Entry System, Toxicity,

## 1. INTRODUCTION
Artificial intelligence is being used commonly nowadays in the process of drug production because of the efficiency with which AI models can handle the data. Now a days, there are studies using variety of neural networks as a way to study the Quantitative Structure–Activity Relationship (QSAR) of a molecule. Artificial Neural Networks are quite efficient in QSAR analysis of a molecule based on descriptors [1]. Due to advancement in the field of science and technology, new techniques have been developed to characterize the physio-chemical properties of molecules. QSAR are being studied using multiple architectures like Convolutional Neural Networks (CNN), Recurrent Neural Networks (RNN), Backpropagation Neural Network (BNN) [2] etc. The most important advantage of Artificial Neural Networks is the efficiency and speed with which it can predict toxicity in early stages of drug development. It is estimated that toxicity is responsible for rejection of one third of drug candidates and hence toxicity in chemicals is a major contributor to the high cost of drug development. Almost all compounds are toxic if taken in overdose and nontoxic if taken in low amount, but the proposed model is using axiom of Paracelcus [3] to predict toxicity at doses that are relevant for a patient during use.

The simplified molecular-input line-entry system (SMILES) is a method of representing elements in the form of a line notation that describes the structure of various chemical molecules using short American Standard Code for Information Interchange (ASCII Notations). Simplified molecular-input line-entry system has already been used in multiple fronts as an alternative to Quantitative Structure–Property Relationships

(QSPR) [4]. Simplified molecular-input line-entry system notations are easily readable by computer and thus provide a good parameter for machine learning models. SMILES provide efficient descriptors for QSAR analysis and thus has been proven as an effective and efficient parameter for QSAR of various chemical species. SMILES is not a data structure like extended connection table and thus is more efficient. SMILES is in form of a language, with simple symbols (atom and bond symbols). SMILES are compact as well because of which they occupy much less space than other representation structures.

## 2. METHODOLOGY
This paper is based on analysis of currently available studies of various Quantitative Structure–Activity Relationship (QSAR) methods of determining toxicity and then comparison with a machine learning model based on SMILES for determination of toxicity.

The AI model is made using Long-short term memory (LSTM) architecture which is an advanced form of recurrent neural network (RNN). We used LSTM as it provides the most efficient correlations between sequential data.

Our model has been trained on data that has been made public free from sites like Kaggle. We have architecture like RNNs and LSTMs to find the relationship between the sequence of the atoms in the molecule to detect if it toxic or not. We could've work on different neural net architecture like CNN's or perception models, but since it is important in our sequential data to find the correlation between the sequence, we have used LSTM. The sequential data have been encoded in arrays of binary digits for our model to work. The encoded input data is passed through the LSTM Layer and finally to a dense network to predict if toxic with a sigmoid function.

## 3. NEURAL NETWORKS USED FOR QSAR ANALYSIS
Various neural networks are used as per the need in QSAR analysis depending on the need. Single-layer neural networks have been used in QSAR modeling for a long time [5] and nowadays its natural to apply multilayer feed-forward networks for predictions as there has been an increase in data available and rise in computation power. Some commonly used architectures in Quantitative Structure–Activity Relationship (QSAR) based analysis are-

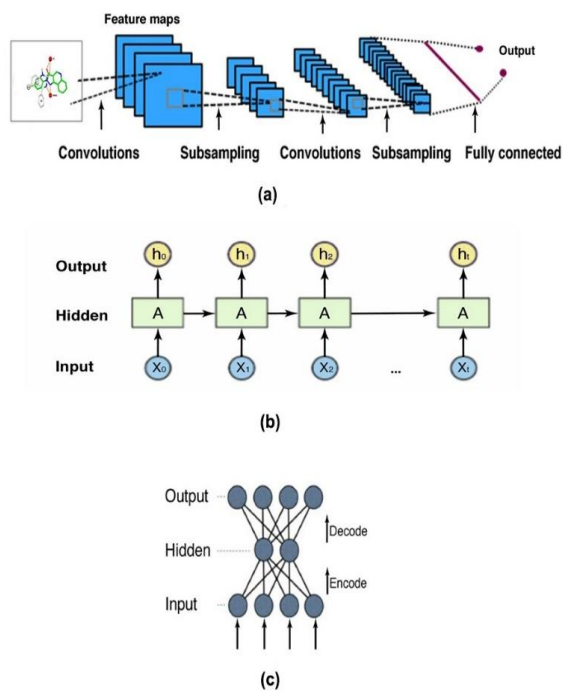

**Fig 1: Neural Networks used in QSAR analysis**

### 3.1 Convolutional Neural Networks (CNN)

CNN (Fig. 1(a)) is widely used for image recognition. It usually contains several convolution layers and subsampling layers. The convolution layer consists of a set of filters (or kernels) which contain receptive fields and learning parameters. CNN largely reduces the number of free parameters learned, thus lowering the consumed memory and increasing the learning speed. Because of high efficiency in image processing and recent advancements in the field of imaging technologies, CNN has become more and more popular in study of chemical species.

### 3.2 Recurrent Neural Networks (RNN)

RNNs (Fig. 1(b)) can take sequential data as input, which is very suitable tasks like language modeling [6]. RNNs are widely used in the field of Natural Language Processing (NLP) and have proven efficient in study of language string notations like SMILES. RNNs have been successful to an extent where they have been used to predict SMILES structure after training on suitable data [7], [8]. RNNs are currently most preferred architecture in the analysis of SMILES notation.

### 3.3 Auto Encoder (AE)

An AE (Fig. 1(c)) is a neural network architecture mainly used for unsupervised learning. It contains an encoder part, which transforms input received into hidden units, and then couples a decoder neural network with the output layer having the same number of nodes as the input layer. Recently, the AE concept has become more popular and more widely used for learning generative models from data. [9].

Deep Neural Networks are quite useful in predicting properties of chemical species because of their ability to analyze multiple descriptors. DNNs are also effective in the process of learning from structures of data directly without using any predefined structure descriptor which makes need of feature selection and reduction procedures unnecessary thus making the prediction efficient and faster.

## 4. SIMPLIFIED MOLECULAR-INPUT LINE-ENTRY SYSTEM

The **simplified molecular-input line-entry system** (**SMILES**) is a method of describing various molecules using American Standard Code for Information Interchange (ASCII) strings. SMILES are useful for studying of chemical properties using computers and software models. SMILES has specified rules for representation of various molecule structure properties. A combination of alphabets inside a square bracket are used to represent atoms ([Al] represents Aluminum). Bonds are reparented by special characters. Single bond is represented by the character '-'. Double, triple, and quadruple bonds are represented by the symbols =, # and $ respectively. '-' can be ignored while representing a single bond. Ring structures are written by breaking each ring at an arbitrary point to make an acyclic structure and adding numerical ring closure labels to show connectivity between non-adjacent atoms. For example, dioxane is written as O1CCOCC1 where 1 represents the cycle present in the molecule. Multiple digits after a single atom indicate multiple ring-closing bonds. For example, an alternative SMILES notation for decalin is C1CCCC2CCCCC12, where the final carbon participates in both ring-closing bonds 1 and 2. SMILES have provision to specify stereochemistry also but it is not compulsory to do so. Configuration around double bonds is specified using the characters / and \ to show directional single bonds adjacent to a double bond. For example, F/C=C/F is one representation of trans-1,2-difluoroethylene.

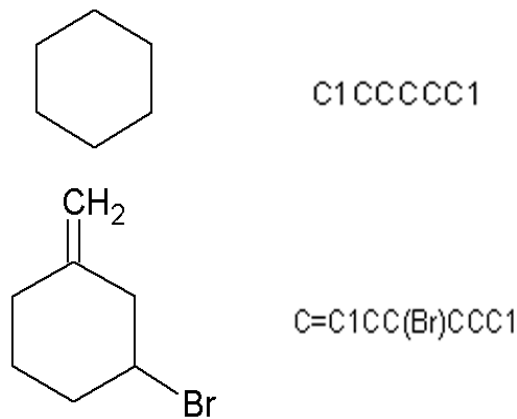

**Fig 2: SMILES Representation**

SMILES have a wide use to make optical descriptors for quantitative structure–activity relationships (QSAR) because of the ease with which ASCII strings can be processed. Continuous work is being done to find possible ways to improve the SMILES based concept of QSPR/QSAR [10]. SMILES is used as a replacement for graph in modelling QSAR/QSPR analysis. SMILES based optical descriptors have been used for modelling HIV-1 PR fullerene-based inhibitors also [11].

## 5. LONG SHORT-TERM MEMORY

**Long Short-Term Memory** (**LSTM**) is an artificial recurrent neural network (RNN) architecture used in the field of deep learning. LSTM is Unlike standard feedforward neural networks which transfer data forward after processing, LSTM has feedback connections in order to store result of current input for use in near future for other predictions. LSTM is capable of predicting from single data (such as images) as well as sequences of data (such as speech or video). For example, LSTM is applicable to tasks such as text recognition, speech recognition etc.

LSTM was designed keeping in mind the need to have a longer memory in order to remember information for longer period of times compared to other deep learning architectures. Structure of LSTM are somewhat similar to logic gates as LSTM have gates to control flow of information. Mainly three gates are used in LSTM which are input gate, output gate and forget gate and all these add up to the efficiency of LSTM in storing feedbacks for a certain period of time. The control of LSTM on selective storing of information which may be useful in future while making a prediction is what makes LSTM efficient while working on sequential data.

SMILES being in form of ASCII sequences, LSTM is suited to make predictions with them. SMILES are sequences of symbols and thus the prediction from each symbol need to be stored for a short term in order to make predictions from the whole sequence as a whole. Thus, in order to make qualitive analysis like determining toxicity of a molecule from SMILES, neural network architecture such as LSTM which have the ability to store results for use in future are efficient.

## 6. EXPERIMENTAL METHODOLOGY

Our approach to solving the sequential nature of the SMILES to be predicted as toxic or not has been solved by the LSTM. The LSTM has 3 main gates; Input, Forget and Output gate which are sigmoid activated, that is the value we get as output from the gates are between 0 and 1. Where '0' signifies blocking and '1' as allowing through the gate.

The equations of the gates are-

$$I(t) = \sigma(Wi[h(t-1), x(t)] + Bi)$$
$$F(t) = \sigma(Wf[h(t-1), x(t)] + Bf)$$
$$O(t) = \sigma(Wo[h(t-1), x(t)] + Bo)$$

Where I(t), F(t), O(t) represents the Input, Forget and Output gate represents Sigmoid activation $W_x$ represent the weights of the different gates(x), h(t-1) represents the output of the previous LSTM block and x(t) represent current inputs and B the corresponding bias of the gates.

The final states and output are represented as

$$\tilde{c}_t = \tanh(w_c[h_{t-1}, x_t] + b_c)$$
$$c_t = f_t * c_{t-1} + i_t * \tilde{c}_t$$
$$h_t = o_t * \tanh(c^t)$$

Where, $c^t$ represents the cell state at time t, and $\tilde{c}^t$ represents candidate for the cell state, and $h^t$ the final output of the LSTM cell.

Figure 3 shows various gates at any given time t. Gates turn to facilitate flow of data from one gate to another as per the output of functions. On giving the values in said equations, gates can be analyzed.

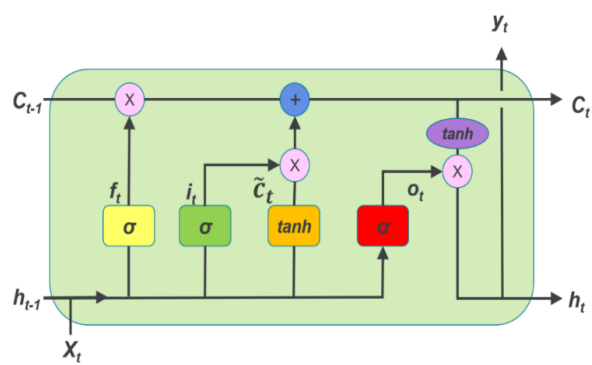

**Fig 3: LSTM at timestep t**

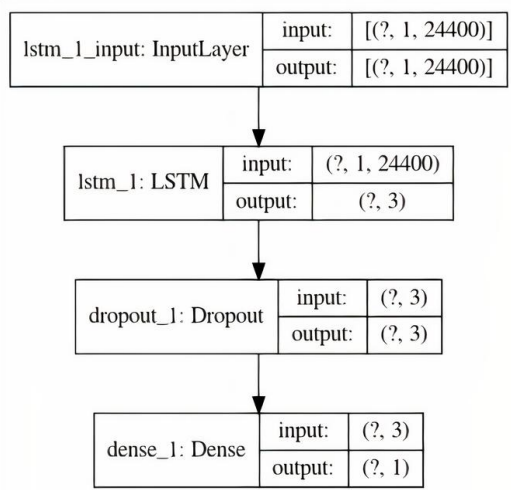

**Fig. 4- Our Model Architecture**

For our model which had sequential data represented as SMILES, we have used one LSTM layer. The maximum length of the Sequence was 400, so we made each combination a length of 400, by padding them with 0's at the end. There was a total of 61 different characters in the SMILE sequences. So, the shape of the dataset became (batch size, 400, 61) where the batch size is defined by us. But we wanted to make our LSTM give one input for every one element of the sequence, so we reshaped the size and it takes the shape (batch size, 1, 24400) since (400*61 = 24400). The drop out layer of probability 0.2 was given so that our model doesn't get overfitted. In the last layer, we only have a single neuron which is Sigmoid activated, that is it gives a number between 0 and 1. We train this whole network with the dataset for 7 epochs and we use a batch size of 32.

## 7. RESULT AND ANALYSIS

The data we are using is highly imbalanced as the number of class '0' is more than twice the number of class '1'. Due to the imbalance, our model may not converge properly and may give unreliable predictions. To make our network work with the data, we applied Undersampling. We first separated the data in Train and Test parts.

In the Training part of the dataset, we have sampled the data randomly and made the data in such a way that we have almost the same number of both classes as this reduced the effect of imbalanced data on training. This process is called Undersampling. After training on this undersampled data we tested our model with the Test Classes which doesn't affect the testing. We tried working with various batch sizes in order to find the optimal result for our data and model. From the experiments with different batch size of the Training data we see that there is an upright change in test parameters like accuracy and Area Under the Curve of Receiver Characteristic Operator (AUROC) with varying batch sizes. We see that although the batch size 32 has the lowest training accuracy compared to other batch sizes, its testing and AUROC is better than that of others. Moreover, we observe that in general increasing the batch size shows increase in training accuracy but the testing accuracy and AUROC almost remains constant. We observe that even though the accuracy keeps changing for different batch sizes other properties remain same and respectable. The results received from the machine learning model are promising and it has the prospect of giving better results with provision of better-balanced data.

**Table 1. Results from our model**

| Batch Size | Training Accuracy | Testing Accuracy | AUROC |
|---|---|---|---|
| 8 | 83.01% | 75.99% | 0.58 |
| 16 | 85.8% | 73.74% | 0.59 |
| **32** | 76.37% | 84.92% | 0.61 |
| 64 | 86.53 | 73.74% | 0.58 |
| 128 | 86.25 | 73.83% | 0.58 |

## 8. CONCLUSION

Need for efficient and comparatively fast method for detection of toxicity of new drug candidates is ever increasing and Simplified Molecular-Input Line-Entry System (SMILES) shows to be a good basis of development of machine learning based models for development of models to solve the problem. Machine learning architectures when provided with correct parameters or data have shown promising results in solving real world problems in a much efficient manner compared to conventional methods.